\documentclass[10pt,letterpaper]{article}
\usepackage[top=0.85in,left=2.75in,footskip=0.75in,marginparwidth=2in]{geometry}

\usepackage[utf8]{inputenc}

\usepackage{cite}

\usepackage{nameref,hyperref}

\usepackage[right]{lineno}

\usepackage{microtype}
\DisableLigatures[f]{encoding = *, family = * }

\raggedright
\usepackage{changepage}

\usepackage[aboveskip=1pt,labelfont=bf,labelsep=period,singlelinecheck=off]{caption}

\makeatletter
\renewcommand{\@biblabel}[1]{\quad#1.}
\makeatother

\usepackage{lastpage,fancyhdr,graphicx}
\usepackage{epstopdf}
\fancyheadoffset[L]{2.25in}
\fancyfootoffset[L]{2.25in}

\usepackage{color}

\definecolor{Gray}{gray}{.25}

\usepackage{graphicx}

\usepackage{sidecap}

\usepackage{wrapfig}
\usepackage[pscoord]{eso-pic}
\usepackage[fulladjust]{marginnote}
\reversemarginpar

\begin{document}
\vspace*{0.35in}

\begin{flushleft}
{\Large
\textbf\newline{Model-based quantitative methods to predict irradiation-induced swelling in alloys}
}
\newline
\\
Wei Ge\textsuperscript{1},
Shijun Zhao\textsuperscript{2},
Chenxu Wang\textsuperscript{1},
Haocheng Liu\textsuperscript{1,4},
Yue Su\textsuperscript{1},
Jia Huang\textsuperscript{1},
Zhiying Gao\textsuperscript{1},
Jianming Xue\textsuperscript{1},
Steven J. Zinkle\textsuperscript{3,*},
Yugang Wang\textsuperscript{1,*}
\\
\bigskip
\bf{1} State Key Laboratory of Nuclear Physics and Technology, Center for Applied Physics and Technology, Peking University, Beijing, 100871, China
\\
\bf{2} Department of Mechanical Engineering, City University of Hong Kong, Hong Kong, China
\\
\bf{3} Department of Nuclear Engineering, University of Tennessee, Knoxville, TN, 37996, USA
\\
\bf{4} State Power Investment Corporation Research Institute, Beijing 102209, China
\\
\bigskip

Wei Ge and Shijun Zhao contribute equally to this paper

* ygwang@pku.edu.cn(Yugang Wang) and szinkle@utk.edu (Steven J. Zinkle)

\end{flushleft}

\section*{Abstract}
Predicting volume swelling of structural materials in nuclear reactors under high-dose neutron irradiations based on existing low-dose experiments or irradiation data with high-dose-rate energetic particles has been a long-standing challenge for safety evaluation and rapidly screening irradiation-resistant materials in nuclear energy systems. Here, we build an Additional Defect Absorption Model that describes the irradiation-induced swelling effects produced by energetic electrons, heavy-ions, and neutrons by considering additional defect sinks inherent in the irradiation process. Based on this model, we establish quantitative methods to predict high-dose swelling from low-dose behavior and obtain the equivalent irradiation dose for different energetic particles when the dose rates differ by several orders of magnitude. Furthermore, we propose a universal parameter to characterize the swelling resistance of various alloys and predict their radiation tolerances under different radiation conditions. This work provides quantitative prediction methods for evaluating irradiation-induced swelling effects of structural materials, which is critical to the safety and material development for advanced nuclear reactors.


\section*{Introduction}

The development of advanced nuclear energy systems is one of the most promising ways to address the anticipated global climate change and energy crisis. In nuclear reactors, the economics and safety depend critically on the durability of structural materials in the reactor core, where the materials need to survive extremely harsh environments, including irradiation by energetic neutrons up to 200 dpa (displacement per atom) \cite{Zinkle2014,Norgett1975}. In addition, recent efforts to extend the lifetime of current nuclear reactors also raise higher demands on irradiation tolerance of structural materials\cite{Zinkle2013}. Therefore, evaluating and quantitatively predicting the performance of materials under high-dose neutron irradiation and R\&D of radiation-resistant materials are critical for the realization of advanced nuclear energy systems and the lifetime extension of current operating reactors. 

Neutron irradiation can lead to severe property deterioration of structural materials, such as volume swelling, hardening, embrittlement, creep, phase instability, etc\cite{Zinkle2012}. Among them, volume swelling is a vital material degradation mechanism. The degree of volume swelling has been used as one of the most critical criteria for materials selection in the reactor design\cite{Mattas1984}, and the swelling phenomena have been widely investigated since the 1960s\cite{Cawthorne1967, Mattas1984}. Although a linear relationship between the swelling rate and the irradiation dose is expected at high swelling levels \cite{Was2007}, it is often difficult to evaluate the swelling trend as it is intrinsically a non-linear relationship at low to intermediate doses. Particularly, swelling in materials exhibits several distinct stages, i.e., a nucleation/incubation stage, a transient stage, and a steady-growth stage. Until now, there is no suitable general model to describe the volume swelling at different doses, which greatly hinders the evaluation of structural materials in nuclear plants and the development of new materials for advanced nuclear reactors. 

Traditionally, neutron radiation effects are studied through test reactors. After irradiation, the bulk properties of irradiated materials are tested, and the microstructures are characterized by post-irradiation examination (PIE) techniques. However, for advanced nuclear reactors, it would take a long time (more than ten years) for irradiation with reactor neutrons because of the high dpa level to achieve lifetime-relevant doses, resulting in an enormous cost and testing difficulty associated with potentially high radioactivation of materials. Irradiation-induced volumetric swelling in alloys is a bulk property of materials that can be measured with TEM (transmission electron microscopy) since the dimensional changes are typically well correlated with void formation\cite{Zinkle1989,Garner2020}. Since the 1970s, the energetic electron-beam from HVEM (high voltage electron microscope) and heavy ions from accelerators have been extensively used for irradiation and swelling investigation based on TEM analysis \cite{El-Atwani2019, Massee2015, Agarwal2020, Du2018, Wang2019, Was2007}. For example, some experts claimed that the void swelling of thin samples irradiated by HVEM could represent the bulk property, and the fitting result showed a swelling rate of (dose)1.58.\cite{Garner1972} Experimental results also show that the swelling rates decrease significantly with the increase of dose rates. Typical examples are that much lower swelling rates for heavy-ion and electron irradiations are found than neutron irradiation due to their higher damage rate\cite{Gigax2015, Mansur1993, Was2015, Jiao2018, Okita2004, Garner2012, Was2007, Zinkle2012}. Indeed, ion- and electron-beam irradiation experiments could achieve high damage levels in a short time and a shallow depth at a low cost with no or little radioactivation of materials\cite{Was2012, Zinkle2017a,Hosemann2012,  Was2015, Zinkle2012,  Zinkle2018}. However, because of the much higher damage rate (103-104 orders of magnitude higher) and the differences in the primary in-cascade process, the extent to which electron or ion irradiation effects can represent neutron irradiation needs to be scrutinized. Notably, there exist two critical and long-standing unsolved questions for irradiation-induced volume swelling in practice: can volume swelling at high doses be predicted from easily accessible low-dose irradiation experiments, and how to quantitatively predict neutron radiation damage from high-dose-rate heavy-ion irradiation experiments? 

Volume swelling originates from the evolution of irradiation-induced defects. It is well established that energetic particles interact with materials by transferring their kinetic energies into the electronic and atomic subsystems of target materials. Irrespective of the dose and dose rate difference, this process can be described by two main simplified processes\cite{Was2007}: (1) Energetic particles first collide with lattice atoms and produce primary knock-on atoms (PKAs), giving rise to Frenkel defects (i.e., vacancies and self-interstitial atoms) with the same overall number of interstitials and vacancies. If the PKAs have sufficient kinetic energy, they can lead to further atomic cascade displacements and produce more point defects and defect clusters. (2) Defect evolution occurs through either recombination of interstitials and vacancies or diffusion-induced defect absorption/accumulation at various defect sinks, such as dislocations, defect clusters, grain boundaries, interfaces, and precipitates. The remaining mobile vacancies aggregate for void formation and growth. For energetic heavy-ions and neutrons, their PKAs can induce displacement cascades. In contrast, electron bombardment creates uniform Frenkel defects without cascades. Although these energetic particles have differences in charge states, mass, and damage rate, similar irradiation effects have been observed, especially the microstructural changes in materials under irradiation such as volume swelling\cite{Was2007,Zinkle2012}.

There have been sustained efforts to understand and describe the irradiation effects at different doses and dose rates, especially regarding volume swelling. In the early 1970s, Brailsford and Mansur\cite{Brailsford1972, Mansur1978} developed a model that states that changes of an irradiation variable can be accommodated by a shift in other variables so as to keep the overall defect aggregation properties (such as swelling) invariant\cite{Mansur1978,Mansur1986, Mansur1978a, Mansur1993, Mansur1994}. Was et al. applied this model successfully in case of irradiation effects induced by protons and neutrons at low doses (less than 5 dpa) and low dose rates (~10-6 dpa/s)\cite{Was2014,Was2002}. At high dose and high dose rate, a direct comparison between neutron and ion or electron irradiation effects becomes difficult, and discrepancies have been found\cite{Sun2015,Gan2001,Hardie2013}. Based on the production bias model (PBM), Golubov and Singh et al. formulated a theoretical description of volume swelling in pure metals and simple alloys, but with a number of complicated parameters, making its application difficult, especially for actual alloys with complex chemical compositions and microstructures \cite{Singh1997,Golubov2001, Golubov2012,Okita2004}. Furthermore, it has been shown consistently from frequently worldwide-used ion-irradiation experiments that swelling rates are lower at higher-dose-rate irradiation\cite{Garner2012, Gigax2015, Okita2004, Was2007} Most recently, Ren et. al. proposed a general radiation polarization theory and a sample spinning strategy technique to mitigate the excess polarization artifact and point-defect imbalance which might be crucial for making the simulation of neutron radiation by ion-beam radiation more realistic\cite{ren2020sample}. Therefore, an application-oriented general model and prediction methods of irradiation swelling in alloys at different irradiation doses and dose rates are highly desired for the developments of advanced nuclear energy systems. 

Generally, volume swelling occurs on a relatively large spatial scale and a long-time scale. Consequently, the details of in-cascade characteristics and the transient defect behavior are not so important when comparing irradiation effects at different dose rates or with different particles, which justifies the application of rate theory models\cite{Was2007} by neglecting the details of the atomic displacement process. In addition, volume swelling provides a direct link between the observed void distributions at the microscopic level and the material response at the macroscopic scale, which enables us to establish equivalent relations between irradiations with different energetic particles. Taking into consideration the importance of volume swelling in nuclear reactors, in this paper, we develop an additional defect absorption model (ADAM) that separates material-related properties and irradiation conditions and can be easily applied to predict volume swelling of alloys at different doses and different dose rates. We first introduce ADAM in the context of electron irradiation, where the produced defects are in the form of uniformly distributed point defects without cascades. After validating this model with experimental data, we extend ADAM to the case of neutron and heavy-ion irradiation by including the energetic displacement cascade as one of the stages in the whole defect evolution. Overall, our model provides a unified description of volume swelling at different irradiation conditions and further, can be used to make quantitative predictions of the swelling level of actual alloys at different doses and dose rates based on available experimental data.

\section*{Results}
\subsection*{Swelling induced by energetic electrons}

To illustrate the radiation damage process produced by energetic electrons, we schematically show the irradiation effects in Figure 1, specifically the void formation, at different dose rates. 

\begin{figure}[!ht]
\includegraphics[width=\textwidth]{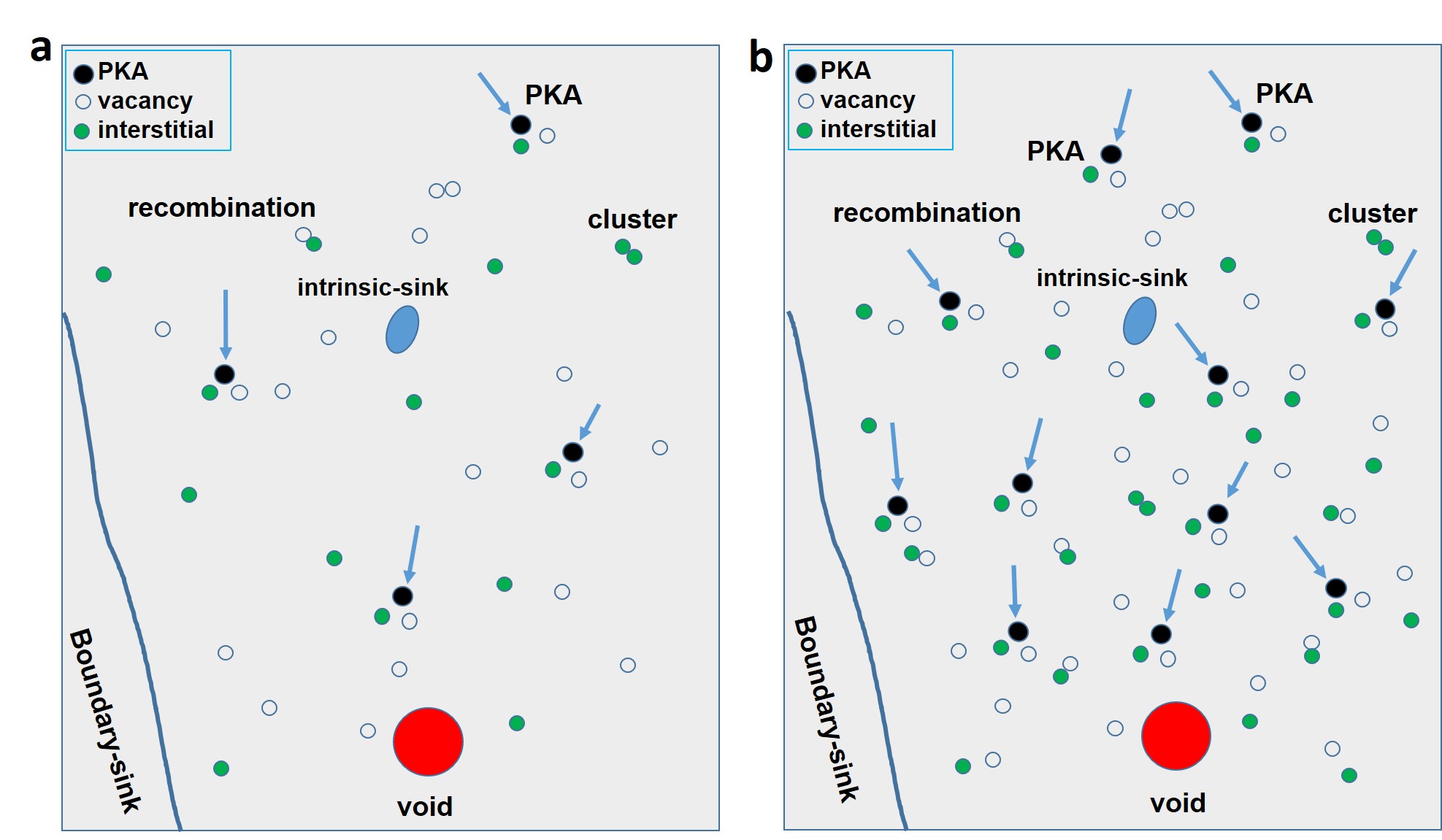}
\caption{\textbf{Schematic illustration of the electron irradiation effects in materials at low and high dose rates}. The energetic electrons generate uniformly distributed point defects. (a) shows the defect process at low dose rates. The defects can either recombine or be absorbed by defect sinks, including dislocations and boundaries. The surviving defects lead to void formation and growth. (b) shows the defect process at high dose rates, where the irradiation induces additional defect sinks can annihilate and absorb mobile defects.}
\label{fig1}
\end{figure}

Electron irradiation induces uniformly distributed point defects. When the irradiation dose rate is high, there is a high probability that Frenkel defects and a few small defect clusters are produced in a high density in a short time period. If the dose rate is sufficiently high, e.g., $10^{-2}-10^{-3}$ dpa/s for electrons in HVEM, the newly produced point defects could be annihilated by the near-by and co-existing Frenkel defects or small clusters. The effect of this process is similar to the capture of defects by the intrinsic sinks in the unirradiated alloys. So, we classify the irradiation-generated near-by and co-existing Frenkel defects or small clusters at high dose rates as additional sinks or irradiation-generated defect sinks. These additional sinks enhance vacancy absorption and reduce the vacancy flux contributing to the void nucleation and growth. Meanwhile, the high-dose-rate irradiation may make the lattice atoms vibrate quickly and increase the local temperature (dynamic temperature), increasing the mobility of defects in a small region and enhancing the additional defect absorption locally. To account for these effects of enhanced defect absorption, we introduce an additional term $S'$ into the rate theory to describe the loss of defects by the irradiation-generated defect sinks. The total adsorption rate then becomes $S_i=S_{i0}+S_i'$ and $S_v=S_{v0}+S_v'$, in which $S_{i0}$ and $S_{v0}$ are the absorption rate of the intrinsic defect sinks of the materials toward interstitials and vacancies respectively, and $S_i'$ and $S_v'$ are the absorption rate of irradiation-generated defect sinks for interstitials and vacancies, respectively. 

The additional sinks mainly result from the product of the near-by and co-existing Frenkel defects at high dose rates, which may not be the final stable extended defects (clusters) but can enhance local defect absorption in a locally “heated” region associated with high dose rates, leading to point defect being captured locally. Under irradiation, the generation rate of such additional defect sinks $\chi_{i,v}$ in reducing mobile defects is proportional to the defect generation rate G for interstitials and vacancies respectively, i.e. $\chi_i=\delta_i G_i$ and $\chi_v=\delta_v G_v$ at relevant temperatures for swelling, where $\delta_{i,v}$ are the ratio constants. Consequently, the net absorption rate of the additional sinks is proportional to their generation rate, i.e. $S_i'=\gamma_i \chi_i=\gamma_i \delta_i G_i$ and  $S_v'=\gamma_v \chi_v=\gamma_v \delta_v G_v$, with $\gamma_{i,v}$ being the proportional constants. Since the electron-irradiation-induced defects are generated in the form of Frenkel pairs, the generation rate of interstitials $G_i$ and the generation rate of vacancies $G_v$ are the same, $G_i=G_v=G$. The additional sink introduced in our model is dynamic, depending on local temperature and nearby defect cluster concentrations, etc. Since the role of the additional sink is the same as the intrinsic sink in capturing interstitials and vacancies, the characteristics of these two types of sinks are similar.  The concentration and the sink strength of the additional sinks may vary during irradiation, but the sink absorption rate bias may not change significantly. Therefore, we assume that the sink bias towards vacancies and interstitials from all sinks is roughly proportional to that of intrinsic sinks, i.e., $S_i/S_v =(S_{i0}+S_i')/(S_{v0}+S_i' )=b, S_{i0}/S_{v0} =b_0$, and $b=\beta b_0$. As a result, the total sink absorption rate for interstitials and vacancies are $S_i=S_{i0}+\gamma_i \delta_i G_i$, and $S_v=S_{v0}+\gamma_v \delta_v G_v$, respectively. Note that additional sinks might be unstable because of the possible irradiation-induced dissociation. In our model, the absorption rate of additional sinks can be taken as the net absorption rate, i.e. $S'=S_{absorption}'-S_{dissociation}'$. Therefore, only those defects that survive from dissociations contribute to defect absorption in our model.

With the additional defect sinks, we solve the classical rate theory equation for void growth. We consider the sink dominant case since the total defect sinks play majority roles in the considered conditions. The void-induced swelling is derived as:

\begin{equation}
    swelling(\%)=\alpha \left[\frac{\Delta_{dpa,ele}}{1+k'G_{dpa,ele}}\right]^{3/2}-c.
\end{equation}

where  $\alpha=\frac{4\pi}{3V_0}\left[2\Omega \frac{D_v b-D_i}{b\eta S_{v0}}\right]^{3/2}, k'=\frac{\gamma_v \delta_v}{\eta S_{v0}}$, and $c=\frac{4\pi r_0^3}{3V_0}$. Here $\eta$ is the ratio of the dose rate ($G_{dpa}, dpa/s$) and defect generation rate (G), $V_0$ is the original volume without voids, $\Omega$ is the defect volume, and $D_{v,i}$ is the diffusion coefficient of vacancies/interstitials. We also introduce $r_0$ to account for the void nucleation stage. Specifically, the value of $r_0$ is the critical size of a void embryo that must be achieved in order for the embryo to grow into a void. Eq. (1) provides the relation between volume swelling and irradiation dose at different dose rates. It indicates that at a fixed dose rate, swelling increases with irradiation dose slowly when the dose is relatively low, but it undergoes a sharp increase at high doses. Furthermore, Eq.(1) predicts lower swelling with increasing dose rate at a given irradiation dose due to extension of the transient regime, which is in accordance with experiments \cite{Garner2012,Okita2007,Okita2004}.

The above model provides the quantitative dependence of volume swelling on different parameters at a fixed temperature, including materials properties ($\alpha$ and c), irradiation dose ($\Delta_{dpa,ele}$), and dose rate ($G_{dpa,ele}$). For materials properties, $\alpha$ is governed by intrinsic properties, such as defect diffusion coefficients and the absorption rate of intrinsic defect sinks, including the influence of temperature, pressure, and cold-work conditions on these properties. The parameter c is governed by the void nucleation ability of materials.

\begin{figure}[!ht]
\includegraphics[width=\textwidth]{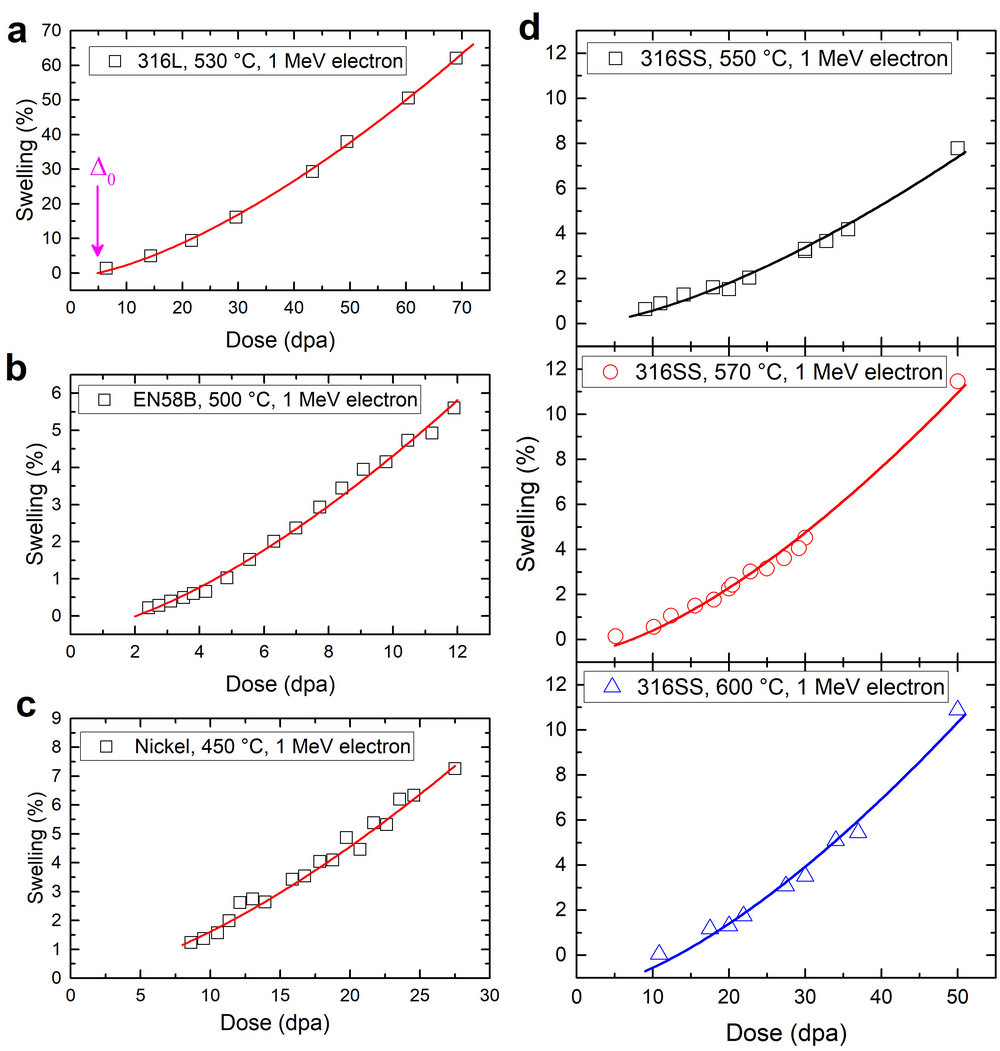}
\caption{\textbf{Validation of the model against volume swelling data from electron irradiation experiments}. (a) 316L irradiated by electrons at 530 °C \cite{Garner1972}, (b) EN58B irradiated by electrons at 500 °C \cite{Norris1971}, (c) nickel irradiated by electrons at 450 °C \cite{Norris1971}, (d) 316SS irradiated by electrons at different temperatures \cite{Hishinuma1977}. The solid lines are our model-fitting results. (Note: some data point with high swelling rate are not included since the saturation may result from surface effects of thin TEM samples)}
\label{fig2}
\end{figure}

In Eq. (1), $k'$ is related to the additional sink absorption rate per defect generation rate, scaled by  $\gamma_{i,v}$, $\delta_{i,v}$ and the absorption rate of intrinsic sinks. Since this model accounts for the additional defect absorption at high dose rates, it is named as Additional Defect Absorption Model. We validate ADAM based on volume swelling data from electron irradiation with HVEM at different conditions. With the irradiation-generated sinks considered, we are able to fit the experimental data to Eq. (1). The results shown in Figure 2 suggest that ADAM agrees well with experimental data, indicating our model provides a robust description of the dose dependence of volume swelling. 

It is shown in Figure 2 that volume swelling may start to grow continuously only after a threshold irradiation dose. This corresponds to the nucleation process or incubation period for volume swelling that is frequently observed in experiments \cite{Garner2000,Garner2012}. By extrapolating to the dose where the swelling equals to zero, we obtain the threshold dose $\Delta_0$ after which pronounced swelling begins (marked with an arrow in Figure 2(a)):

\begin{equation}
    \Delta_0=\frac{r_0^2}{2\Omega} \frac{(1+ \frac{\gamma_v \delta_v}{\eta S_{v0}} G_{dpa} )b\eta S_{v0}}{D_v b-D_i}
\end{equation}

which reasonably relates to $r_0$ and other materials parameters. Besides, $\Delta_0$ increases linearly with increasing $G_{dpa}$ , a signature of the dose rate effects on the nucleation/incubation regime. Our model thus provides a general description of the swelling over the whole dose range.

In Figure 2, all the irradiation experiments were conducted at a fixed dose rate (constant $G_{dpa,ele}$). As a consequence, we cannot isolate the parameter $k'$ but can only take  $k' G_{dpa,ele}$ as a whole in our model fitting. Therefore, there are only two parameters to obtain the relationship between the swelling and dose, i.e., $\alpha[1/(1+k' G_{dpa,ele} )]^{3/2}$ and c. The obtained parameters are provided in the Supplementary materials. It is found that c is small in all cases, suggesting a short incubation period for void swelling in electron irradiation. The prefactor $\alpha[1/(1+k' G_{dpa,ele} )]^{3/2}$ increases with increasing temperature for 316SS, which agrees with the higher swelling observed in experiments at 600 ℃.

\subsection*{Swelling induced by energetic heavy-ions and neutrons}

In contrast to the isolated Frenkel pairs produced by electron irradiation, irradiation by heavy-ions and neutrons induces energetic displacement cascades with nearly the same features at the early stage of defect production. The damage process created by heavy-ions and neutrons at different dose rates is schematically illustrated in Figure 3. After the PKAs and then the collision cascades are produced by heavy-ions and neutrons, athermal recombination between vacancies and interstitials dominates the in-cascade process \cite{Nordlund2018}. We consider that this process is similar to the situation of electron irradiation at a higher beam current, and the same idea of additional sink is used to account for the enhanced defect absorption. Although the defects induced by electron and ion/neutron irradiations may be different, all these irradiation sources produce defect clusters. Particularly, due to the high dose rate ($10^{-3}-10^{-2}$ dpa/s) of high voltage electrons, the produced Frenkel defects are densely distributed spatially, forming high-density defect spots. Such high-density defects are highly likely to aggregate to defect clusters in their evolution processes. For ion/neutron irradiations that create densely populated cascades and subcascades, defect clusters are directly formed. In this sense, defect clusters can be created in both electron and ion/neutron irradiations, contributing to subsequent defect evolution. The fate of these defect clusters depends on various factors, such as irradiation-induced dissolution. The in-cascade process in the case of neutron or heavy-ion irradiation may affect the number of residual interstitials and vacancies entering into the governing equations \cite{Golubov2000,Golubov2012}. The afterward process of defect evolution under these conditions is similar to that of electron irradiations in terms of volume swelling. By treating the in-cascade process as one of the stages of whole defect evolution (since the volume swelling occurs both at a large spatial scale and a long-time scale), we introduce the irradiation-generated defect sinks as stated above. Some experimental and simulation results \cite{Kiritani1994,Pokor2004,Stoller2008,Muroga1990,Phythian1995,Gan2001} showed that the neutron irradiation-induced stable dislocation loop or cluster density increase quickly with dose and then saturate at around 2-3 dpa. So, at the high dose range, these stable and extended defects as sinks to absorb mobile defects could be classified into the intrinsic defect sinks, i.e., $S_{v0}$  and $S_{i0}$. Comparing to neutron irradiation, the dose rate of heavy-ion irradiation is $10^3-10^4$ higher, similar to electron irradiation, and will also increase the local dynamic temperature. As the energy of majority PKAs produced by neutron irradiation in iron is below 10 keV \cite{gilbert2015energy}, we calculate a single cascade morphology, as shown in Figure 4(a) ,  produced by 10 keV PKA(10 keV Fe ion) through SRIM program \cite{ziegler2010srim} in \textit{Full Damage Cascades} mode. And because the mean free path of neutron is very large (about 4 centimeters in iron), neutron-induced PKAs are spatially separated. Also with SRIM calculation, the cascades produced by PKAs along the transport track of incident 1 MeV Fe ion are drawn in Figure 4(b) and Figure 4(c) . Unlike neutron irradiation, a single heavy ion can induce dense PKAs along its transport track. The cascades from these PKAs are produced in picoseconds \cite{antoshchenkova2015fragmentation} and are close to each other. Thus, during heavy-ion irradiations, the additional sinks from the near-by and co-existing defects, some of which are transitional defects, could also enhance local defect absorption at a high dynamic temperature, leading to the extra capture of defects in a local region that survive the in-cascade process. Recent MD results for high energy PKAs also demonstrated that there are strong interactions between the closed-by subcascades, which would help annihilating pre-created defects.\cite{de2018model}

\begin{figure}[!ht]
\includegraphics[width=\textwidth]{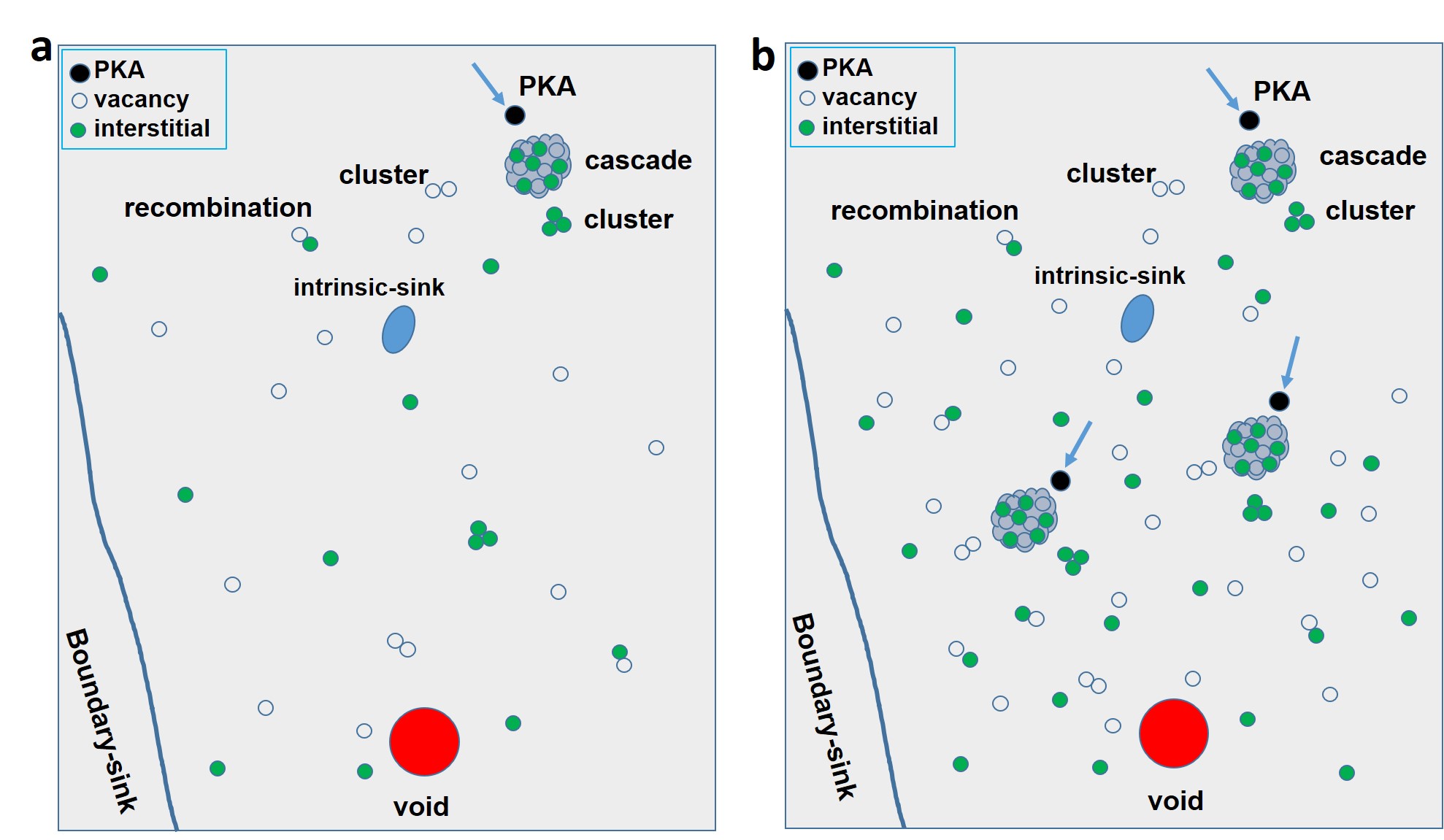}
\caption{\textbf{Schematic illustration of the heavy-ion and neutron irradiation effects in materials at low and high dose rates}. The PKAs from heavy-ion and neutrons produce defects through displacement cascades. (a) shows the defect process at low dose rates, and (b) shows the defect process at high dose rates.}
\label{fig3}
\end{figure}

\begin{figure}[!ht]
\includegraphics[width=\textwidth]{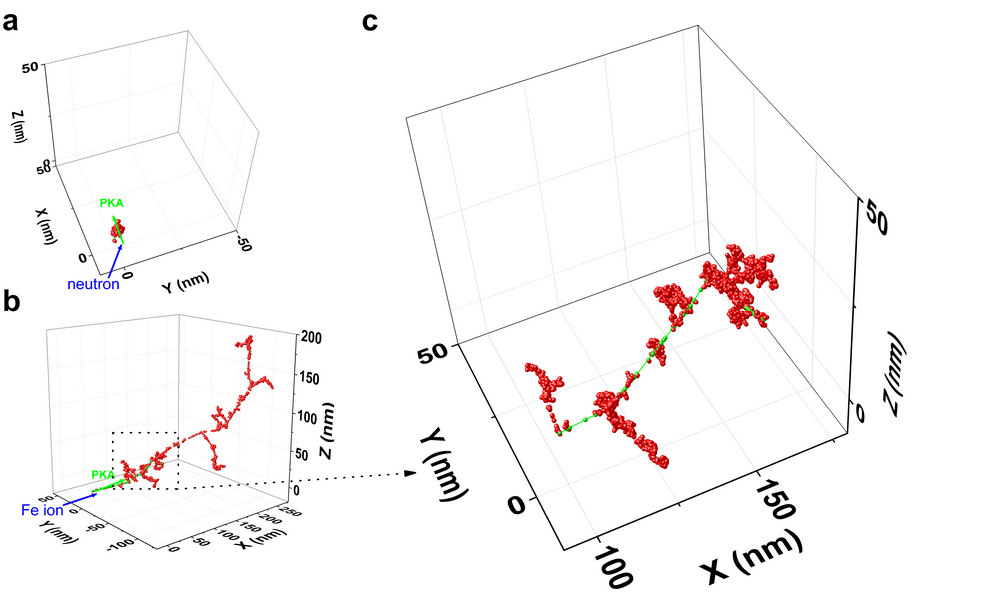}
\caption{\textbf{cascade morphology produced by neutron or ion}. (a) a cascade produced by 10 keV PKA from neutron irradiation in iron; (b) spatially close cascades produced by PKAs from 1 MeV $\textrm{Fe}^{2+}$ irradiation in iron; (c) the partial enlarged view of (b).}
\label{fig4}
\end{figure}

With the above consideration, we apply the derived ADAM to the case of heavy-ion and neutron irradiations. The volume swelling becomes

\begin{equation}
    swelling(\%)=\alpha \left[\frac{\Delta_{dpa}}{1+k'G_{dpa}} \right]^{3/2} -c
\end{equation}

Here we have omitted the specific denotation in $\Delta_{dpa}$ and $G_{dpa}$  for different energetic particles, which means this relation is suitable for all these energetic particles. Based on this equation, we fit experimental data collected from experiments for different alloys at varying temperatures and for the same alloy at varying dose rates. As representative cases, Figure 5 shows the fitting results for irradiation data in different material conditions, including varying materials, temperatures, pressure, and helium incorporation (different $\alpha$ , $k'$, and c characterizing material conditions), while Figure 6 demonstrates the fitting results for the same material irradiated at different dose rates (different $G_{dpa}$  but with same $\alpha$ , $k'$, and c) \cite{Garner1983a,Zinkle2013a,Getto2016, Dubuisson1992,Garner2000,Garner2012,Holmes1969,Norris1971,Okita2004,Sun2015,Zinkle2013}. (A more thorough validation of ADAM is provided in Supplement material). The fitting results shown in Figure 5 and Figure 6 demonstrate ADAM is suitable for describing different alloys at neutron and heavy-ion irradiation conditions in terms of volume swelling at the same elevated temperature or in a narrow temperature range. Particularly, our model captures the swelling trend even when He is present in the material, as shown in Figure 5 (e) and (f).

\begin{figure}[!ht]
\includegraphics[width=\textwidth]{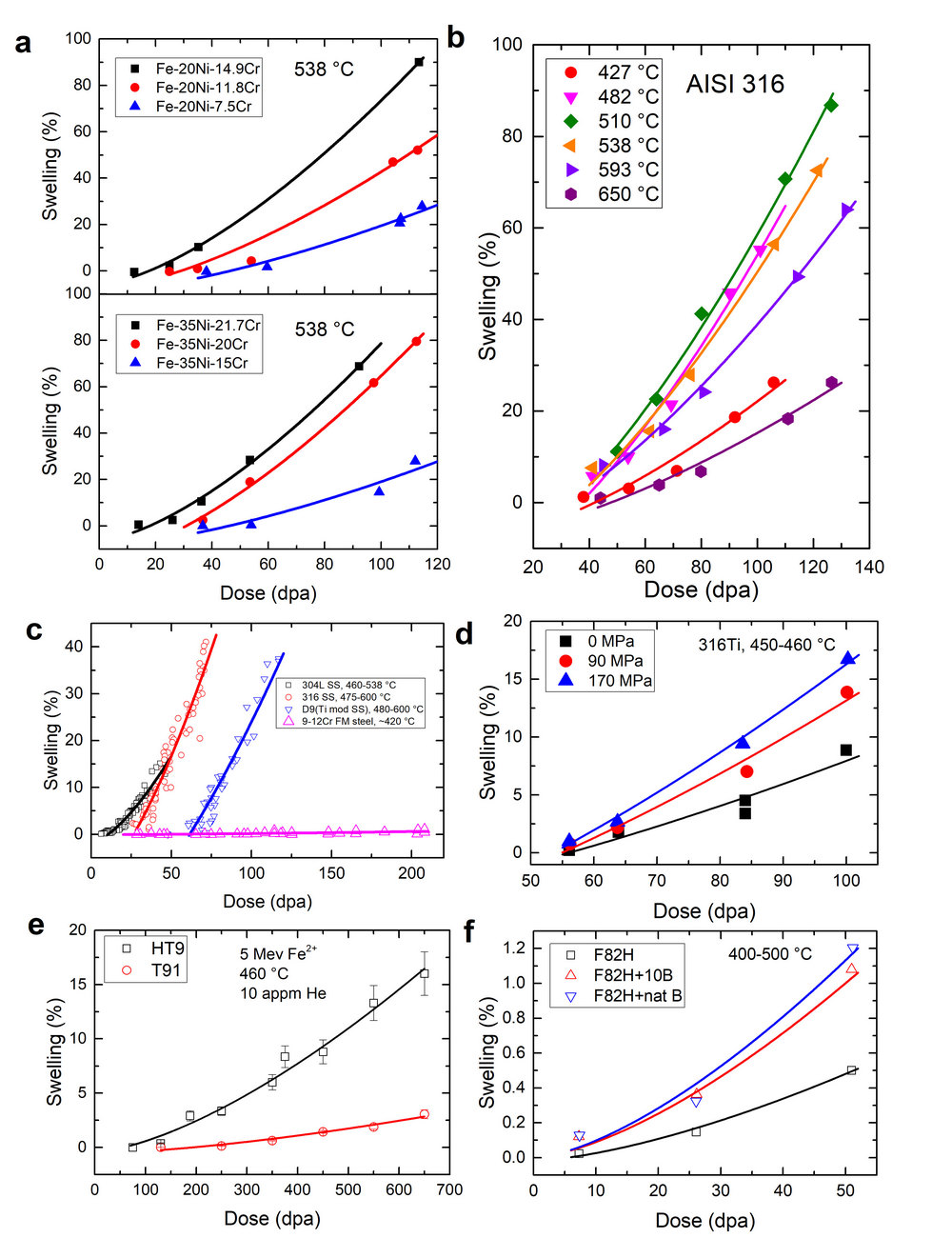}
\caption{\textbf{Validation of the derived model using available irradiation data on different alloys or alloy conditions (different $\alpha$, $k'$, and c)}. (a) Fe-20Ni-xCr and Fe-35Ni-xCr irradiated by neutrons at 538 ℃ \cite{Garner2000}; (b) AISI316 irradiated by neutrons at different temperatures \cite{Garner2000}; (c) 304L SS, 316SS, D9, and 9-12Cr FM irradiated by neutrons \cite{Garner2000,Garner1983a,Zinkle2013}; (d) 316Ti SS irradiated by neutrons at different pressure \cite{Dubuisson1992}; (e) HT9 and T91 irradiated by 5 MeV Fe++ at 460 ℃ \cite{Getto2016}; (f) F82H irradiated by neutrons with different helium injection \cite{Zinkle2013a}. The solid lines are our model-fitting results.}
\label{fig5}
\end{figure}

\subsection*{Prediction methods based on ADAM}

For fast screening and evaluation of structural materials for advanced reactors and current reactor lifetime extension, we predict volume swelling in alloys from low-dose irradiation experiments to high-dose cases based on ADAM. Note that such high-dose data usually requires at least several years to carry out neutron irradiation experiments and PIE. The fitting results displayed in Figure 2, Figure 5, and Figure 6 provide the growth trend of swelling at high doses beyond those in the experimental data, representing the prediction merit of ADAM. In addition, the fitting results in Figure 6 indicate the quantitative correlation of the swelling rate with irradiation dose at different dose rates ($G_{dpa}$ ), or with different energetic particles, making the equivalence possible between different energetic particles, especially between heavy-ion and neutron irradiations in term of volume swelling in the same material at elevated temperatures.
When we compare two irradiation conditions that lead to different swelling levels in the same material and the same elevated temperature, we have

\begin{equation}
    \frac{swelling_1}{swelling_2}=\frac{\alpha \left[\frac{\Delta_{dpa1}}{1+k'G_{dpa1}} \right]^{3/2} -c}{\alpha \left[\frac{\Delta_{dpa2}}{1+k'G_{dpa2}} \right]^{3/2} -c}
\end{equation}

In this case, derivation of the irradiation conditions with a given swelling level based on another irradiation experiment would require the knowledge of $\alpha$ , $k'$, and c. When the swelling level is the same for the identical material, an equivalent relation between the irradiation dose and dose rate can be obtained as:

\begin{equation}
    \frac{\Delta_{dpa1}}{1+k'G_{dpa1}}=\frac{\Delta_{dpa2}}{1+k'G_{dpa2}}
\end{equation}

which is only governed by the parameter of $k'$.

\begin{figure}[!ht]
\includegraphics[width=\textwidth]{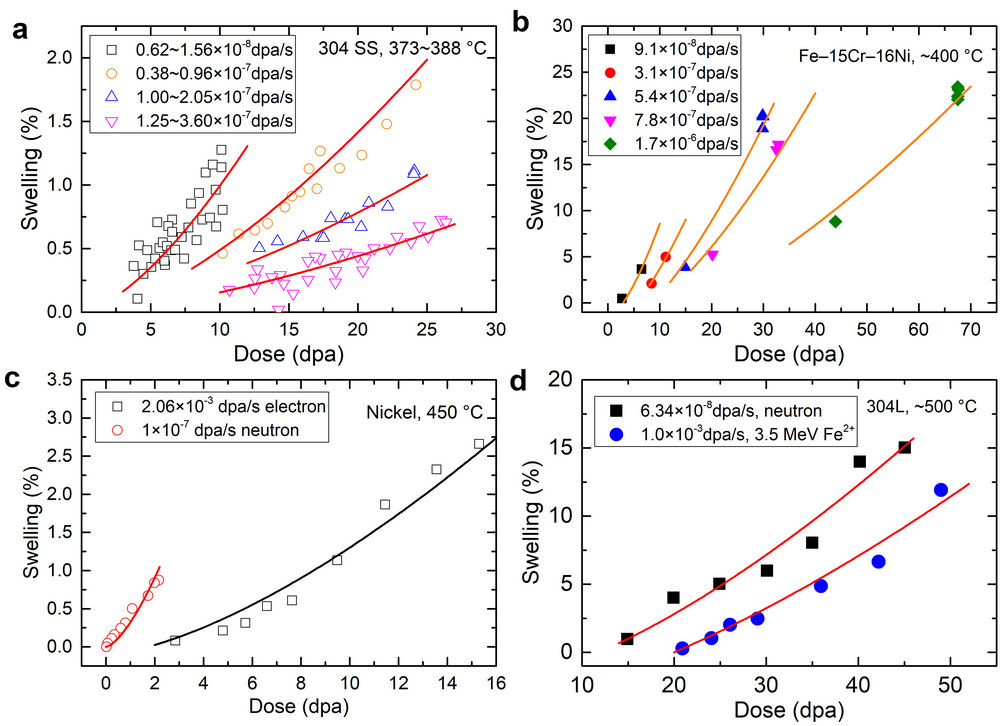}
\caption{\textbf{Validation of the derived model using available irradiation data with different dose rates for the same alloy (constant $\alpha$ , $k'$, and c)}.  (a) 304 SS irradiated with different dose-rate neutrons \cite{Garner2012}, (b) Fe-15Cr-16Ni irradiated with different dose-rate neutrons \cite{Okita2004}, (c) Ni irradiated by 1 MeV electrons and neutrons\cite{Holmes1969,Norris1971}, and (d) 304L SS irradiated by neutrons and heavy ions\cite{Sun2015}. The solid lines are our model-fitting results.}
\label{fig6}
\end{figure}

Equations (4-5) serve as the basis for predicting the swelling level for an identical material at the same or similar elevated temperature at given irradiation conditions, as well as the equivalent irradiation conditions at the same swelling level based on available experiments. In Figure 7, we show three examples of such predictions. In Figure 7(a), the parameters $\alpha$ , $k'$, and c are calculated from the low-dose irradiation data at 435‒487 °C. Then Eq. (5) is used to predict the swelling rate at higher doses. Our prediction is consistent with another set of high-dose irradiation data at 446 °C. Figure 7(b) and (c) illustrate the derivation of equivalent irradiation conditions with the same swelling level, which only requires $k'$. In Figure 7(b), we calculate $k'$ for Fe-15Cr-16Ni-0.25Ti from three sets of irradiation data (solid black symbols)\cite{Garner2012}. Based on $k'$, the irradiation dose required to produce the same swelling at $7.8\times 10^{-7}$ dpa/s can be deduced from the data at $1.7\times 10^{-6}$ dpa/s. The predicted data (empty pentagons) lie within the range of experimental data. In Figure 7(c), the equivalent neutron irradiation dose is predicted from the self-ion (Fe ions) irradiation experiment, with the $k'$ parameter calculated from two sets of irradiation data at different dose rates (solid black symbols)\cite{Getto2016,VanDenBosch2013}. It can be seen that the prediction agrees reasonably well with available neutron irradiation data. Therefore, these examples suggest that our derived additional defect absorption model is capable of predicting volume swelling at high doses from the low dose data, correlating the irradiation dose with different given dose rates, and quantitatively predicting equivalent dose from high dose-rate experiments (such as heavy-ions) to low dose-rate irradiation (such as neutrons).

\begin{figure}[!ht]
\includegraphics[width=0.6\textwidth]{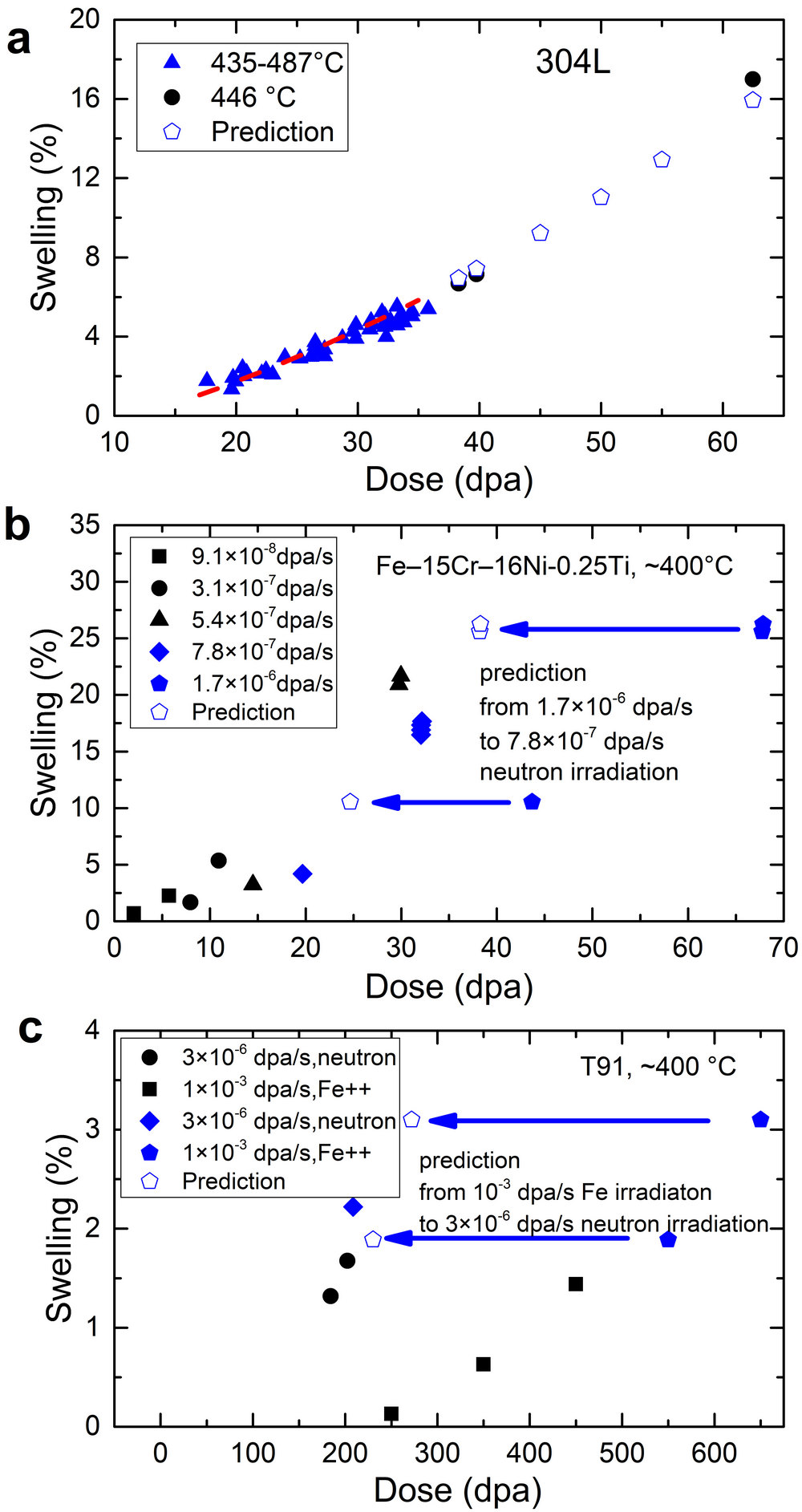}
\caption{\textbf{Quantitative prediction of swelling level by ADAM}. (a) Prediction of swelling level in 304L from low doses to high doses. The parameters are calculated from a set of irradiation data (blue triangle symbols) \cite{Garner1983a}. (b) Prediction for Fe-15Cr-16Ni-0.25Ti from $1.7\times10^{-6}$ dpa/s to $7.8\times10^{-7}$ dpa/s at the same swelling level. The parameters are calculated from three sets of irradiation data (solid black symbols) \cite{Garner2012}. (c) Prediction for T91 from self-ion irradiation ($1\times10^{-3}$ dpa/s) to neutron irradiation ($3\times10^{-6}$ dpa/s) at the same swelling level. The parameters are calculated from two sets of irradiation data with different dose rates (solid black symbols) \cite{Getto2016,VanDenBosch2013}}
\label{fig7}
\end{figure}

\subsection*{Parameters in ADAM governing swelling}

Irradiation effects in materials are complex processes, including defects production and evolution. Previous works have established that volumetric swelling can be influenced by different factors, including properties of materials, e.g., compositions and their atomic weight, crystal structures, grain boundaries and interfaces, atomic bonds, solute segregation and precipitation, the diffusivity of defects, sink strength, etc., at a fixed temperature, and irradiation conditions, i.e., PKA spectrum, irradiation temperature, dose and dose rate\cite{Okita2007,Okita2004}. Based on ADAM, Eq. (3) shows that irradiation-induced swelling is governed by $\alpha$ , $k'$, c, and $\Delta_{dpa}$. The parameter $\alpha$ contains different intrinsic properties of materials (e.g., diffusivities of vacancies and interstitials at fixed temperatures, various intrinsic sink strength) and determines how fast the swelling grows. So, a higher $\alpha$ means faster swelling growth. The value of c is also a material property and relates to the nucleation stage or incubation period for void evolution, which is governed by the threshold dose $\Delta_0$ due to the nucleation/incubation process for swelling. Therefore, a larger $\Delta_0$ indicates a longer nucleation/incubation time, suggesting better swelling resistance. The value of $\Delta_0$ is very important for the development of nuclear materials since the extension of the incubation period is the only meaningful solution for designing swelling-resistant materials up to high doses. The overall swelling of materials relies on both $\Delta_0$ and $\alpha$ since the dose-dependent swelling includes an initial nucleation/incubation period, followed by a growth stage \cite{Wolfer1984,Russell1971}. Finally, $k'$ relates to irradiation conditions and characterizes the ability of materials to absorb mobile defects by the irradiation-generated additional sinks in response to the high dose rate irradiations, which is related to $\gamma_{i,v}$, $\delta_{i,v}$, and the absorption rate of the intrinsic defect sinks. Since irradiation-induced defects, which are the source of additional defect absorption, strongly depend on crystal structures and microstructures, the parameter of $\delta_v$ is expected to be relating to these materials conditions. Therefore, $k'$ is related to both intrinsic sinks and additional sink generation processes, indirectly related to crystal structure and microstructures. The purpose of $k'$ in our model is to reconcile the high density of additional sinks generated at high dose rates, which tends to suppress swelling. As such, it is a key parameter to reveal the impact of dose rate on swelling, as encountered in the comparison between ion and neutron irradiations. All these parameters from our model are provided in the Supplement materials. In this work, we did not explicitly consider the temperature. Therefore, our model describes swelling from different irradiation conditions at the same temperature, or in a narrow temperature window. Since material properties change with temperatures, the swelling also depends on temperature besides the above three parameters. The relations of swelling with all these different parameters are provided in Figure 8.

It is well known that the swelling rate of some alloys exhibits a peak at an intermediate temperature, as shown in the upper panel of Figure 8(a). At low temperatures, defect mobility is low, and a relatively high density of irradiation-induced defect sinks are produced, both of which limits void growth. When the temperature is sufficiently high, swelling is also reduced since the mobility of interstitials and vacancies is high, and the recombination between them increases significantly. Besides, vacancy supersaturation is reduced compared to thermal equilibrium, and the emission of vacancies from cavities increases. The obtained parameters $\alpha$ and $\Delta_0$ from our model based on the experimental data agree with this conclusion. In particular, $\alpha$ exhibits a peak, whereas $\Delta_0$ shows a dip around the peak swelling temperature. 

The dependence of $\alpha$ and $\Delta_0$ on alloy composition is shown in Figure 8(b) for several well-studied alloys. In both Fe-35Ni-xCr and Fe-20Ni-xCr, $\alpha$ decreases, whereas $\Delta_0$ increases with decreasing Cr content. Therefore, the results suggest that reducing Cr enhances the swelling resistance for these alloys. The variation of $\Delta_0$ is more significant than $\alpha$ , signifying a primary influence of Cr on the duration of the incubation regime \cite{Garner2000}. This agrees with prior phenomenological experimental assessments of Fe-Ni-Cr austenitic alloys that the chemical changes predominantly affect the incubation dose rather than the steady-state swelling rate \cite{Garner2012,Okita2005}. To access the parameter of $k'$ for different materials, we need experimental data from different dose rates. Among all the data that we have gathered, the swelling data on Fe-15Cr-16Ni and Fe-15Cr-16Ni-0.25Ti can be used to compare $k'$ for these two materials, as they were irradiated with varying dose and dose rate at the same temperature. Figure 8(c) displays the dependence of $k'$ and $\Delta_0$ on the dose rate for Fe-15Cr-16Ni and Fe-15Cr-16Ni-0.25Ti. For the same material, it is seen that $k'$ is constant. It’s interesting to find that Fe-15Cr-16Ni-0.25Ti, with better swelling resistance than Fe-15Cr-16Ni \cite{Garner2000}, has a smaller $k'$, which indicates that if an alloy possesses more intrinsic vacancy absorption centers, or higher intrinsic sink strength, i.e., larger $S_{v0}$, the irradiation-induced additional sinks are less important for volume swelling. On the other hand, $\Delta_0$ increases linearly with the dose rate for both alloys, consistent with Eq. (2). This conclusion is in line with experiments \cite{Okita2005}.

\begin{figure}[!ht]
\includegraphics[width=\textwidth]{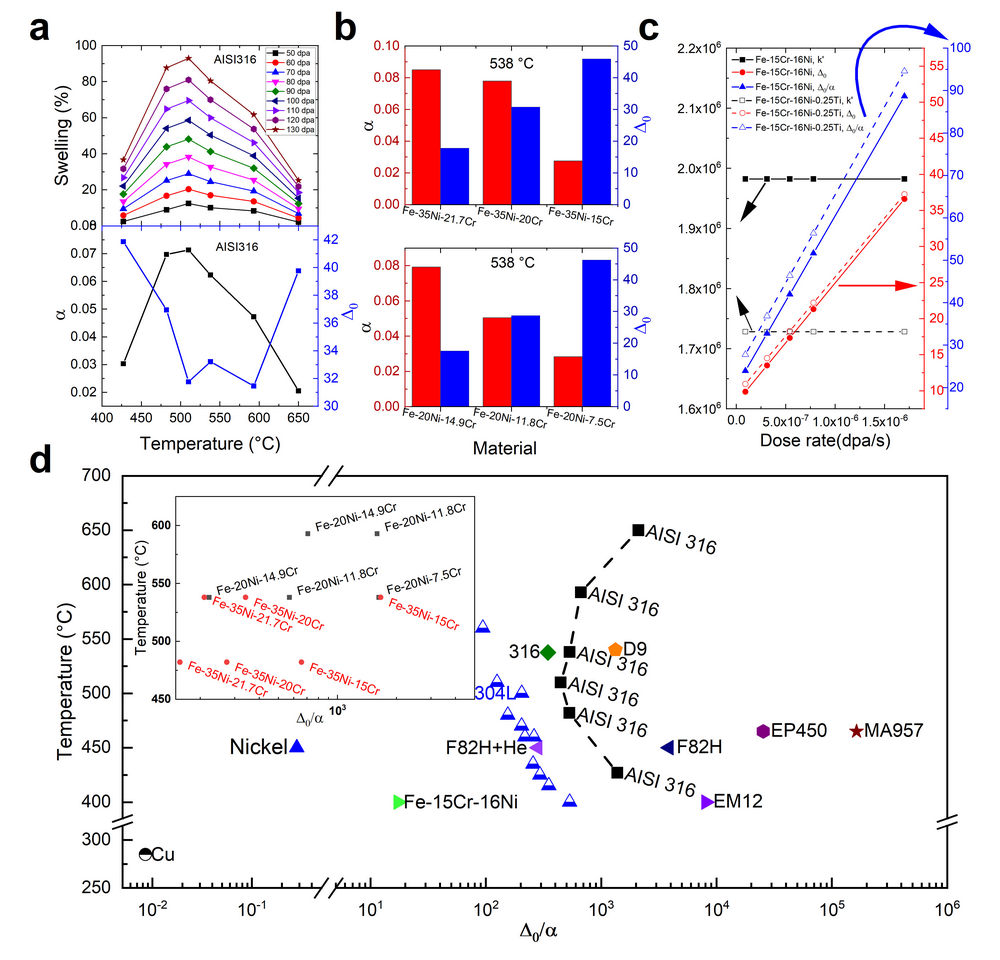}
\caption{\textbf{Dependence of swelling (characterized by $\Delta_0$ , $\alpha$ , and $k'$) on different factors}. (a) Temperature \cite{Garner2000}, (b) materials \cite{Garner2000}, and (c) dose rate \cite{Garner2012,Okita2005}. A summary of the results provided in (d) shows the relations of swelling (characterized by $\Delta_0/\alpha$) with temperature for different materials under neutron irradiation \cite{zinkle2017,Dubuisson1993,Garner2000,Garner1983a,Holmes1969,Okita2004,Singh1993,Sun2015,Zinkle2013a,Zinkle2013}, where the irradiation dose rates are comparable.}
\label{fig8}
\end{figure}

Based on the above discussion, a material with enhanced swelling resistance would have a high $\Delta_0$ but a low $\alpha$ so that there are a prolonged incubation period and a low growth rate. To quantify this relation, we use the simple parameter ratio of $\Delta_0/\alpha$ as a figure of merit to characterize the relative swelling resistance of materials at comparable dose rates. With this definition, larger values of $\Delta_0/\alpha$ correspond with better swelling resistance in a material. A summary of $\Delta_0/\alpha$ values for different materials under neutron irradiation is demonstrated in Figure 8(d), together with the temperature at which $\Delta_0/\alpha$ is obtained. Among all the materials, Cu exhibits the lowest $\Delta_0/\alpha$ due to its poor swelling resistance. On the other hand, MA957, a type of ODS (Oxide Dispersion Strengthened) steel, possesses the highest $\Delta_0/\alpha$ among the materials examined, suggesting that its excellent ability to suppress swelling is due to the high density of nano-particles that provide numerous defect sinks to absorb the mobile vacancies. For the same material AISI316, the curve of $\Delta_0/\alpha$ with the temperature at reactor-relevant dose rates is in accordance with the experimental swelling peak at about 510°C. Note that $\Delta_0/\alpha$ also depends on the dose rate, as shown in Figure 8(c); the comparison in Figure 8(d) is made at comparable dose rates. It also can be seen from Figure 8(c) that a smaller $k'$ corresponds to a larger $\Delta_0/\alpha$ at a given dose rate, indicating that an alloy with higher intrinsic sink strength exhibits better swelling resistance. 

Helium, which can be produced through transmutation reactions under fast fission and fusion neutron irradiation, is of great concern for nuclear materials. The results shown in Figure 5(e) and (f) suggest that ADAM can be used to describe the effects of He on swelling. As the migration of He to sinks, e.g. grain boundaries and dislocations, will reduce sink capture efficiency \cite{Dai2020,li2019radiation,farrell1980experimental}, thus reducing sink absorption rate $S_{v0}$, increasing parameter $\alpha=\frac{4\pi}{3V_0}\left[2\Omega \frac{D_v b-D_i}{b\eta S_{v0}}\right]^{3/2}$, reducing threshold dose  $\Delta_0=\frac{r_0^2}{2\Omega} \frac{(1+ \frac{\gamma_v \delta_v}{\eta S_{v0}} G_{dpa} )b\eta S_{v0}}{D_v b-D_i}$, and facilitating swelling in materials (as indicated by the decrease of $\Delta_0/\alpha$ for “F82H” to “F82H+He” as shown in Figure 8(d)).

\subsection*{Implication of the prediction neutron irradiation from heavy-ion irradiation }

By introducing the irradiation-generated additional defect sinks, we provide a simple unified model (ADAM) to describe the volume swelling in alloys irradiated by different energetic particles with varying dose rates at a given temperature. These additional sinks come from the near-by and co-existing defects at high dose rates, enhancing local defect absorption, compensating for the effects of high dose rates, and leading to fewer mobile defects that contribute to volume swelling. This is consistent with the general observation that the efficiency of producing swelling is lower for high-dose-rate than low-dose-rate irradiations (shown in Figure 6), which highlights the role of the additional sinks in influencing volume swelling when the dose rates vary by several orders (e.g., heavy-ions vs. neutrons). To provide further validation of our model, we have simulated void swelling in the framework of phase-field modeling, as provided in the Supplement materials. The modeling results are consistent with our ADAM analysis, where swelling grows steadily with increasing doses. Inspiringly, the swelling relation with irradiation dose can be well modeled by our ADAM formulae. Besides, the phase-field results reproduce the suppressed swelling trend with an increasing dose rate, in line with our ADAM prediction and experiments. Therefore, these integrated computational efforts shed light on the swelling mechanism in materials under different irradiation conditions, which serve as important guidelines for designing swelling-resistant structural materials. 

At the same swelling level, our model indicates that by selecting suitable experimental conditions, heavy-ions, especially self-ion irradiation, can be used not only to investigate the irradiation damage mechanism but also to predict equivalent neutron irradiation dose at the same volume swelling level. This prediction from ion irradiation to neutron radiation makes heavy-ion irradiation quite important and useful in the fast screening of advanced nuclear structural materials. However, the high dose rate screening results need to be adjusted by a factor to account for dose rate effects. From Eq. (5), we have the factor  $\Delta_{ions}/\Delta_{neutrons} =(1+k' G_{ions})/(1+k' G_{neutrons} )>1$, where $\Delta_{ions}$ and $\Delta_{neutrons}$  are doses at the same swelling level for ions and neutrons, respectively. As a consequence, the difference between the dose of heavy-ions and neutrons $(\Delta_{ions}-\Delta_{neutrons})=\Delta_{ions} k'  (G_{ions}-G_{neutrons})/(1+k' G_{ions} ))$ increases with increasing ion dose ($\Delta_{ions}$). Therefore, the radiation tolerance of alloys determined solely by heavy-ion irradiation results may be quantitatively overestimated by this factor, e.g., 1.38 for 304L in Figure 6(d) and 2.39 for T91 in Figure 7(c), where the differences in factors are due to different material parameters of $k'$. It means that a dose of 300 dpa or 600 dpa of ion irradiation of T91 is only equivalent to 125 dpa or 250 dpa (5/12) of neutron irradiation in terms of void swelling, respectively. Our model indicates that in the high dose regime, increasing the irradiation dose of heavy-ions only leads to a limited increase in the equivalent neutron dose. 

In our model, we assume that there is no saturation of void swelling in materials under irradiation, agreeing with most experiments. In principle, when voids are large enough, the void surfaces may become the predominant sink for all types of radiation defects. Besides, there is a possibility of the reduction of network dislocation densities at high doses. In these cases, void swelling may saturate at high irradiation doses. However, most experiments show no saturation of void swelling within a given irradiation dose. Particularly, only under certain conditions is found saturation of swelling. For instance, electron irradiation of stainless-steel samples that are too thin leads to an apparent saturation of swelling, which is actually an artifact of the surface impact. The irradiation of bulk samples, on the other hand, gives rise to steady swelling without saturation. To validate our model, we have chosen the experimental data that mostly show no saturation of void swelling under irradiation. Accordingly, the total sink absorption rate bias (k) is constant in our model. Therefore, our model cannot be used to predict swelling in the saturation regime. Overall, experimental evidence for a high-dose swelling saturation regime is very limited, and convincing evidence for swelling saturation did not emerge until ~100 dpa for neutron-irradiated Fe-Cr-Ni alloys\cite{Garner2000} or ~1000 dpa for ion irradiated T91.\cite{gigax2016radiation} Our model is proved to be applicable at such high doses.

Our additional defect absorption model is application-oriented and is based on a simplified description of irradiation-induced volume swelling in alloys considering point defect production and evolution, such as those in the electron irradiation. We extended ADAM to heavy-ion and neutron irradiation cases, but the detailed in-cascade atomistic processes are not explicitly considered. We also did not consider the details of sink strength from different types of defect sinks since our focus is on the effects induced by the dose and dose rate difference. Besides, we assume that the strength of intrinsic defect sinks is approximately constant during the irradiation. The details of defect forms, such as different types of defect clusters, are also neglected. Therefore, although our model can be used to describe the whole stage of volume swelling and fit the experimental data well at different irradiation conditions, the quantitative prediction of equivalent irradiation dose at different dose rates and from low-dose to high-dose cases may not be used for all irradiation conditions. In ADAM, we have omitted the contribution of thermal vacancies at the void surface. This approximation is reasonable at intermediate temperatures but may not be suitable at high temperatures (such as $T > 0.5 T_{Melting}$). Finally, it’s worth noting that it is impractical to build equivalent relations based on the present model for all the irradiation effects produced by neutrons and heavy-ions, such as irradiation-induced hardness, embrittlement and creep. Hence, there is still a high challenge to study and quantify comprehensive materials’ performance under high-dose neutron irradiation by virtue of heavy-ion or self-ion irradiation techniques.

In summary, considering the main characteristics of defect evolution under energetic particle irradiation, this paper constructs an integrated kinetic rate theory model to correlate the volume swelling in alloys under different irradiation conditions. Specifically, our model introduces an additional defect sink in the defect master equations when considering the differences in dose rates of irradiations. With the irradiation-generated additional defect sink, our model provides a unified description of volume swelling at different doses and dose rates, especially for different energetic particles. Based on ADAM, quantitative methods are established to predict irradiation swelling in alloys for the advanced nuclear energy system, from low doses to high doses, from one dose rate to another dose rate. It can particularly be used for predicting the equivalent dose from heavy-ion irradiation to neutron irradiation experiments. In addition, the parameters of $\Delta_0$ and $\alpha$ in ADAM well capture the swelling features, and the combined parameter of $\Delta_0/\alpha$ is proposed to quantitatively characterize the relative swelling resistance of alloys for nuclear plants. Therefore, this work provides an innovative practical solution to evaluate the swelling effects of the structural materials in reactor cores and a fast screening way to assess the performance of newly-developed alloys for advanced nuclear energy systems.

\section*{Materials and Methods}

The governing equations for interstitials and vacancies respectively under electron irradiation, are

\begin{equation}
    \frac{\partial c_i}{\partial \tau}=G_i-R_{iv}c_i c_v-S_i c_i,
\end{equation}

\begin{equation}
    \frac{\partial c_v}{\partial \tau}=G_v-R_{iv}c_i c_v-S_v c_v.
\end{equation}

In these two equations, $c_i$ and $c_v$ are the concentration of interstitials and vacancies respectively, and $R_{iv}$ is the recombination rate of interstitials and vacancies. If local equilibrium is reached ($\tau$ is sufficiently large), meaning steady-state conditions, and assuming $G_i=G_v=G$, the solutions of Eqs. (6) and (7) give the defect concentrations:

\begin{equation}
    c_i=\frac{S_v}{S_i} c_v, c_v=\frac{2G}{S_v+\sqrt{4GR_{iv} \frac{S_v}{S_i}+S_v^2}}
\end{equation}

The steady-state solutions are used in the following volume swelling model since it is generally considered that defects can achieve their equilibrated states quickly compared to the time scale of volume growth \cite{Was2007}. The classical rate theory for void growth is \cite{Was2007}:

\begin{equation}
    \frac{dr}{dt}=\frac{\Omega}{r}[D_v c_v-D_i c_i]
\end{equation}

where $r$ is the void radius, and $\Omega$ is the defect volume. Here we have omitted the concentrations of thermal vacancies in the void surface. After inserting the solutions of $c_{v,i}$, we have

\begin{equation}
    \frac{dr}{dt}=\frac{\Omega}{r}[D_v S_i-D_i S_v]\frac{-1+\sqrt{1+\frac{4GR_{iv}}{S_i S_v}}}{2R_{iv}}
\end{equation}

We consider the sink dominant case, i.e.,$\frac{4GR_{iv}}{S_i S_v}\ll 1$. In this case, Eq. (10) can be simplified to:

\begin{equation}
    \frac{dr}{dt}=\frac{\Omega}{r}[D_v S_i-D_i S_v]\frac{G}{S_i S_v}
\end{equation}

The solution of this equation is $r=\left[t \frac{2G\Omega}{S_{v0}+\gamma_v \delta_v G} (D_v-\frac{D_i}{b}) \right]^{1/2}$. After inserting the assumption $b=\beta b_0$, the total void-induced swelling can be written as:

\begin{equation}
    swelling(\%)=\frac{\Delta V}{V_0}=\frac{\frac{4\pi r^3}{3}-\frac{4\pi r_0^3}{3}}{V_0}=\frac{4\pi}{3V_0}\left[ t \frac{2G\Omega}{S_{v0}+\gamma_v \delta_v G} \frac{D_v\beta S_{i0}-D_i S_{v0}}{\beta S_{i0} } \right]^{3/2}-\frac{4\pi r_0^3}{3V_0}
\end{equation}

Considering the relations among dose ($\Delta_{dpa}$, dpa), dose rate ($G_{dpa}$, dpa/s), and defect generation rate (G) of electrons, $G_{dpa,ele}=\eta G$, and $\Delta_{dpa,ele}=G_{dpa,ele} t$, we can rewrite the swelling to

\begin{equation}
    swelling(\%)
    =\frac{4\pi}{3V_0}\left[ 2\Omega \frac{D_v \beta S_{i0} -D_i S_{v0} }{\beta \eta S_{i0} S_{v0} } \frac{\Delta_{dpa}}{1+\frac{\gamma_v \delta_v}{\eta S_{v0}} G_{dpa}} \right]^{3/2}-\frac{4\pi r_0^3}{3V_0}
    =\alpha \left[ \frac{\Delta_{dpa,ele}}{1+k'G_{dpa,ele}}\right]^{3/2}-c.     
\end{equation}


\section*{Data Availability}
All data generated or analyzed during this study are included in this published article (and its supplementary information files).

\section*{Code Availability}
The Phase-Field Modeling program in this paper are deposited at a website, https://github.com/pku-ion-beam/PFM.



\section*{Acknowledgments}
The authors thank Dr. Roger Stoller and Dr. Yong Dai for valuable discussions and suggestions. This work was supported by National Science Foundation of China (Grant No. 11935004 and Grant No. 12192280). S. Zhao was support by City University of Hong Kong (Grant No. 9610425).

\section*{Contributions}
Y. W. and S. J. Z. conceived the research. W.G. and S.Z. performed the theoretical investigation. C. W., H. L., Y.S, and J. X. assisted in data collection and analysis. S. Z., Y.W., W.G., C. W. and S. J. Z. prepared the manuscript. All authors discussed the results, commented on the manuscript, and contributed to the writing of the paper.


\bibliography{main}

\begin{thebibliography}{10}

\bibitem{Zinkle2014}
S.J. Zinkle and L.L. Snead.
\newblock {Designing Radiation Resistance in Materials for Fusion Energy}.
\newblock {\em Annual Review of Materials Research}, 44(1):241--267, jul 2014.

\bibitem{Norgett1975}
MJ~Norgett, MT~Robinson, and IM~Torrens.
\newblock {A proposed method of calculating displacement dose rates}.
\newblock {\em Nuclear Engineering and Design}, 33(1):50--54, 1975.

\bibitem{Zinkle2013}
S.~J. Zinkle and G.~S. Was.
\newblock {Materials challenges in nuclear energy}.
\newblock {\em Acta Materialia}, 61(3):735--758, feb 2013.

\bibitem{Zinkle2012}
S.J. Zinkle.
\newblock {Radiation-Induced Effects on Microstructure}.
\newblock In Rudy J~M Konings and Roger~E Stoller, editors, {\em Comprehensive
  Nuclear Materials}, volume~1, pages 65--98. Elsevier, 2012.

\bibitem{Mattas1984}
R.~F. Mattas, F.~A. Garner, M.~L. Grossbeck, P.~J. Maziasz, G.~R. Odette, and
  R.~E. Stoller.
\newblock {The impact of swelling on fusion reactor first wall lifetime}.
\newblock {\em Journal of Nuclear Materials}, 122(1-3):230--235, may 1984.

\bibitem{Cawthorne1967}
C.~Cawthorne and E.~J. Fulton.
\newblock {Voids in irradiated stainless steel}, nov 1967.

\bibitem{Was2007}
Gary~S Was.
\newblock {\em {Fundamentals of radiation materials science: metals and
  alloys}}.
\newblock Springer Science \& Business Media, 2007.

\bibitem{Zinkle1989}
S.~J. Zinkle and K.~Farrell.
\newblock {Void swelling and defect cluster formation in reactor-irradiated
  copper}.
\newblock {\em Journal of Nuclear Materials}, 168(3):262--267, dec 1989.

\bibitem{Garner2020}
Frank~A. Garner.
\newblock {Radiation-Induced Damage in Austenitic Structural Steels Used in
  Nuclear Reactors}.
\newblock In Rudy J~M Konings and Roger~E Stoller, editors, {\em Comprehensive
  Nuclear Materials}, pages 57--168. Elsevier, Oxford, second edi edition,
  2020.

\bibitem{El-Atwani2019}
O.~El-Atwani, N.~Li, M.~Li, A.~Devaraj, J.~K.S. Baldwin, M.~M. Schneider,
  D.~Sobieraj, J.~S. Wr{\'{o}}bel, D.~Nguyen-Manh, S.~A. Maloy, and
  E.~Martinez.
\newblock {Outstanding radiation resistance of tungsten-based high-entropy
  alloys}.
\newblock {\em Science Advances}, 5(3):eaav2002, mar 2019.

\bibitem{Massee2015}
Freek Massee, Peter~Oliver Sprau, Yong-Lei Wang, J.~C.~S{\'{e}}amus Davis,
  Gianluca Ghigo, Genda~D. Gu, and Wai-Kwong Kwok.
\newblock {Imaging atomic-scale effects of high-energy ion irradiation on
  superconductivity and vortex pinning in Fe(Se,Te)}.
\newblock {\em Science Advances}, 1(4):e1500033, may 2015.

\bibitem{Agarwal2020}
S.~Agarwal, M.~O. Liedke, A.~C.~L. Jones, E.~Reed, A.~A. Kohnert, B.~P.
  Uberuaga, Y.~Q. Wang, J.~Cooper, D.~Kaoumi, N.~Li, R.~Auguste, P.~Hosemann,
  L.~Capolungo, D.~J. Edwards, M.~Butterling, E.~Hirschmann, A.~Wagner, and
  F.~A. Selim.
\newblock {A new mechanism for void-cascade interaction from nondestructive
  depth-resolved atomic-scale measurements of ion irradiation–induced defects
  in Fe}.
\newblock {\em Science Advances}, 6(31):eaba8437, jul 2020.

\bibitem{Du2018}
Congcong Du, Shenbao Jin, Yuan Fang, Jin Li, Shenyang Hu, Tingting Yang, Ying
  Zhang, Jianyu Huang, Gang Sha, Yugang Wang, Zhongxia Shang, Xinghang Zhang,
  Baoru Sun, Shengwei Xin, and Tongde Shen.
\newblock {Ultrastrong nanocrystalline steel with exceptional thermal stability
  and radiation tolerance}.
\newblock {\em Nature Communications}, 9(1):5389, dec 2018.

\bibitem{Wang2019}
Chenxu Wang, Tengfei Yang, Cameron~L. Tracy, Chenyang Lu, Hui Zhang, Yong-Jie
  Hu, Lumin Wang, Liang Qi, Lin Gu, Qing Huang, Jie Zhang, Jingyang Wang,
  Jianming Xue, Rodney~C. Ewing, and Yugang Wang.
\newblock {Disorder in Mn+1AXn phases at the atomic scale}.
\newblock {\em Nature Communications}, 10(1):622, dec 2019.

\bibitem{Garner1972}
F.~A. Garner and L.~E. Thomas.
\newblock {Production of Voids in Stainless Steel By High-Voltage Electrons.}
\newblock In {\em ASTM Special Technical Publication}, pages 303--323, 100 Barr
  Harbor Drive, 1972. ASTM International.

\bibitem{Gigax2015}
Jonathan~G. Gigax, Eda Aydogan, Tianyi Chen, Di~Chen, Lin Shao, Y.~Wu, W.~Y.
  Lo, Y.~Yang, and F.~A. Garner.
\newblock {The influence of ion beam rastering on the swelling of self-ion
  irradiated pure iron at 450 °C}.
\newblock {\em Journal of Nuclear Materials}, 465:343--348, oct 2015.

\bibitem{Mansur1993}
L.K. Mansur.
\newblock {Theory of transitions in dose dependence of radiation effects in
  structural alloys}.
\newblock {\em Journal of Nuclear Materials}, 206(2-3):306--323, nov 1993.

\bibitem{Was2015}
Gary~S. Was.
\newblock {Challenges to the use of ion irradiation for emulating reactor
  irradiation}.
\newblock {\em Journal of Materials Research}, 30(9):1158--1182, may 2015.

\bibitem{Jiao2018}
Z.~Jiao, J.~Michalicka, and G.~S. Was.
\newblock {Self-ion emulation of high dose neutron irradiated microstructure in
  stainless steels}.
\newblock {\em Journal of Nuclear Materials}, 501:312--318, apr 2018.

\bibitem{Okita2004}
T.~Okita and W.~G. Wolfer.
\newblock {A critical test of the classical rate theory for void swelling}.
\newblock {\em Journal of Nuclear Materials}, 327(2-3):130--139, may 2004.

\bibitem{Garner2012}
F.A. Garner.
\newblock {Radiation Damage in Austenitic Steels}.
\newblock In Rudy J~M Konings and Roger~E Stoller, editors, {\em Comprehensive
  Nuclear Materials}, volume~4, pages 33--95. Elsevier, 2012.

\bibitem{Was2012}
G.S. Was and R.S. Averback.
\newblock {Radiation Damage Using Ion Beams}.
\newblock In Rudy J~M Konings and Roger~E Stoller, editors, {\em Comprehensive
  Nuclear Materials}, volume~1, pages 195--221. Elsevier, jan 2012.

\bibitem{Zinkle2017a}
S.~J. Zinkle.
\newblock {Advanced irradiation-resistant materials for Generation IV nuclear
  reactors}.
\newblock In Pascal Yvon, editor, {\em Structural Materials for Generation IV
  Nuclear Reactors}, pages 569--594. Woodhead Publishing, jan 2017.

\bibitem{Hosemann2012}
Peter Hosemann, Daniel Kiener, Yongqiang Wang, and Stuart~A. Maloy.
\newblock {Issues to consider using nano indentation on shallow ion beam
  irradiated materials}.
\newblock {\em Journal of Nuclear Materials}, 425(1-3):136--139, jun 2012.

\bibitem{Zinkle2018}
S.~J. Zinkle and L.~L. Snead.
\newblock {Opportunities and limitations for ion beams in radiation effects
  studies: Bridging critical gaps between charged particle and neutron
  irradiations}.
\newblock {\em Scripta Materialia}, 143:154--160, jan 2018.

\bibitem{Brailsford1972}
A.~D. Brailsford and R.~Bullough.
\newblock {The rate theory of swelling due to void growth in irradiated
  metals}.
\newblock {\em Journal of Nuclear Materials}, 44(2):121--135, 1972.

\bibitem{Mansur1978}
L.~K. Mansur.
\newblock {Void swelling in metals and alloys under irradiation: An assessment
  of the theory}.
\newblock {\em Nuclear Technology}, 40(1):5--34, 1978.

\bibitem{Mansur1986}
L.~K. Mansur, E.~H. Lee, P.~J. Maziasz, and A.~P. Rowcliffe.
\newblock {Control of helium effects in irradiated materials based on theory
  and experiment}.
\newblock {\em Journal of Nuclear Materials}, 141-143(PART 2):633--646, nov
  1986.

\bibitem{Mansur1978a}
L.~K. Mansur.
\newblock {Correlation of neutron and heavy-ion damage. II. The predicted
  temperature shift if swelling with changes in radiation dose rate}.
\newblock {\em Journal of Nuclear Materials}, 78(1):156--160, nov 1978.

\bibitem{Mansur1994}
L.K. Mansur.
\newblock {Theory and experimental background on dimensional changes in
  irradiated alloys}.
\newblock {\em Journal of Nuclear Materials}, 216(C):97--123, oct 1994.

\bibitem{Was2014}
G.~S. Was, Z.~Jiao, E.~Getto, K.~Sun, A.~M. Monterrosa, S.~A. Maloy,
  O.~Anderoglu, B.~H. Sencer, and M.~Hackett.
\newblock {Emulation of reactor irradiation damage using ion beams}.
\newblock {\em Scripta Materialia}, 88:33--36, oct 2014.

\bibitem{Was2002}
G.~S. Was, J.~T. Busby, T.~Allen, E.~A. Kenik, A.~Jensson, S.~M. Bruemmer,
  J.~Gan, A.~D. Edwards, P.~M. Scott, and P.~L. Andreson.
\newblock {Emulation of neutron irradiation effects with protons: Validation of
  principle}.
\newblock {\em Journal of Nuclear Materials}, 300(2-3):198--216, feb 2002.

\bibitem{Sun2015}
C.~Sun, S.~Zheng, C.~C. Wei, Y.~Wu, L.~Shao, Y.~Yang, K.~T. Hartwig, S.~A.
  Maloy, S.~J. Zinkle, T.~R. Allen, H.~Wang, and X.~Zhang.
\newblock {Superior radiation-resistant nanoengineered austenitic 304L
  stainless steel for applications in extreme radiation environments}.
\newblock {\em Scientific Reports}, 5(1):1--7, jan 2015.

\bibitem{Gan2001}
J.~Gan and G.~S. Was.
\newblock {Microstructure evolution in austenitic Fe-Cr-Ni alloys irradiated
  with rotons: Comparison with neutron-irradiated microstructures}.
\newblock {\em Journal of Nuclear Materials}, 297(2):161--175, aug 2001.

\bibitem{Hardie2013}
Christopher~D. Hardie, Ceri~A. Williams, Shuo Xu, and Steve~G. Roberts.
\newblock {Effects of irradiation temperature and dose rate on the mechanical
  properties of self-ion implanted Fe and Fe-Cr alloys}.
\newblock {\em Journal of Nuclear Materials}, 439(1-3):33--40, aug 2013.

\bibitem{Singh1997}
B.~N. Singh, S.~I. Golubov, H.~Trinkaus, A.~Serra, Yu~N. Osetsky, and A.~V.
  Barashev.
\newblock {Aspects of microstructure evolution under cascade damage
  conditions}.
\newblock {\em Journal of Nuclear Materials}, 251:107--122, nov 1997.

\bibitem{Golubov2001}
S.~I. Golubov, B.~N. Singh, and H.~Trinkaus.
\newblock {On recoil-energy-dependent defect accumulation in pure copper Part
  II. Theoretical treatment}.
\newblock {\em Philosophical Magazine A: Physics of Condensed Matter,
  Structure, Defects and Mechanical Properties}, 81(10):2533--2552, 2001.

\bibitem{Golubov2012}
S.~I. Golubov, A.~V. Barashev, and R.~E. Stoller.
\newblock {Radiation damage theory}.
\newblock In Rudy J~M Konings and Roger~E Stoller, editors, {\em Comprehensive
  Nuclear Materials}, volume~1, pages 357--391. Elsevier Ltd, 2012.

\bibitem{ren2020sample}
Cui-Lan Ren, Yang Yang, Yong-Gang Li, Ping Huai, Zhi-Yuan Zhu, and Ju~Li.
\newblock Sample spinning to mitigate polarization artifact and
  interstitial-vacancy imbalance in ion-beam irradiation.
\newblock {\em npj Computational Materials}, 6(1):1--11, 2020.

\bibitem{Okita2007}
Taira Okita, Toshihiko Sato, Naoto Sekimura, Takeo Iwai, and Francis~A Garner.
\newblock The synergistic influence of temperature and displacement rate on
  microstructural evolution of ion-irradiated fe--15cr--16ni model austenitic
  alloy.
\newblock {\em Journal of nuclear materials}, 367:930--934, 2007.

\bibitem{Norris1971}
D.~I.R. Norris.
\newblock {The use of the high voltage electron microscope to simulate fast
  neutron-induced void swelling in metals}.
\newblock {\em Journal of Nuclear Materials}, 40(1):66--76, jul 1971.

\bibitem{Hishinuma1977}
Akimichi Hishinuma, Yoshio Katano, and Kensuke Shiraishi.
\newblock {Dose and temperature dependence of void swelling in electron
  irradiated stainless steel}.
\newblock {\em Journal of Nuclear Science and Technology}, 14(10):723--730,
  1977.

\bibitem{Garner2000}
F.~A. Garner, M.~B. Toloczko, and B.~H. Sencer.
\newblock {Comparison of swelling and irradiation creep behavior of
  fcc-austenitic and bcc-ferritic/martensitic alloys at high neutron exposure}.
\newblock {\em Journal of Nuclear Materials}, 276(1):123--142, jan 2000.

\bibitem{Nordlund2018}
Kai Nordlund, Steven~J. Zinkle, Andrea~E. Sand, Fredric Granberg, Robert~S.
  Averback, Roger Stoller, Tomoaki Suzudo, Lorenzo Malerba, Florian Banhart,
  William~J. Weber, Francois Willaime, Sergei~L. Dudarev, and David Simeone.
\newblock {Improving atomic displacement and replacement calculations with
  physically realistic damage models}.
\newblock {\em Nature Communications}, 9(1):1084, dec 2018.

\bibitem{Golubov2000}
S.~I. Golubov, B.~N. Singh, and H.~Trinkaus.
\newblock {Defect accumulation in fcc and bcc metals and alloys under cascade
  damage conditions - towards a generalization of the production bias model}.
\newblock {\em Journal of Nuclear Materials}, 276(1):78--89, jan 2000.

\bibitem{Kiritani1994}
M.~Kiritani.
\newblock {Microstructure evolution during irradiation}.
\newblock {\em Journal of Nuclear Materials}, 216(C):220--264, oct 1994.

\bibitem{Pokor2004}
C.~Pokor, Y.~Brechet, P.~Dubuisson, J.~P. Massoud, and A.~Barbu.
\newblock {Irradiation damage in 304 and 316 stainless steels: Experimental
  investigation and modeling. Part I: Evolution of the microstructure}.
\newblock {\em Journal of Nuclear Materials}, 326(1):19--29, mar 2004.

\bibitem{Stoller2008}
Roger~E Stoller and G~Robert Odette.
\newblock A composite model of microstructural evolution in austenitic
  stainless steel under fast neutron irradiation.
\newblock In {\em Radiation-Induced Changes in Microstructure: 13th
  International Symposium (Part I)}. ASTM International, 1987.

\bibitem{Muroga1990}
T.~Muroga, H.~Watanabe, and N.~Yoshida.
\newblock {Correlation of fast neutron, fusion neutron and electron
  irradiations based on the dislocation loop density}.
\newblock {\em Journal of Nuclear Materials}, 174(2-3):282--288, nov 1990.

\bibitem{Phythian1995}
W.~J. Phythian, R.~E. Stoller, A.~J.E. Foreman, A.~F. Calder, and D.~J. Bacon.
\newblock {A comparison of displacement cascades in copper and iron by
  molecular dynamics and its application to microstructural evolution}.
\newblock {\em Journal of Nuclear Materials}, 223(3):245--261, jun 1995.

\bibitem{gilbert2015energy}
Mark~R Gilbert, Jaime Marian, and J-Ch Sublet.
\newblock Energy spectra of primary knock-on atoms under neutron irradiation.
\newblock {\em Journal of nuclear materials}, 467:121--134, 2015.

\bibitem{ziegler2010srim}
James~F Ziegler, Matthias~D Ziegler, and Jochen~P Biersack.
\newblock Srim--the stopping and range of ions in matter (2010).
\newblock {\em Nuclear Instruments and Methods in Physics Research Section B:
  Beam Interactions with Materials and Atoms}, 268(11-12):1818--1823, 2010.

\bibitem{antoshchenkova2015fragmentation}
Ekaterina Antoshchenkova, Laurence Luneville, David Simeone, Roger~E Stoller,
  and Marc Hayoun.
\newblock Fragmentation of displacement cascades into subcascades: A molecular
  dynamics study.
\newblock {\em Journal of Nuclear Materials}, 458:168--175, 2015.

\bibitem{de2018model}
A~De~Backer, Christophe Domain, CS~Becquart, L~Luneville, D~Simeone, Andrea~E
  Sand, and Kai Nordlund.
\newblock A model of defect cluster creation in fragmented cascades in metals
  based on morphological analysis.
\newblock {\em Journal of Physics: Condensed Matter}, 30(40):405701, 2018.

\bibitem{Garner1983a}
F.~A. Garner and D.~L. Porter.
\newblock {Reassessment of the Swelling Behavior of Aisi 304 Stainless Steel.}
\newblock In {\em Proceedings of the International Conference on Dimensional
  Stability and Mechanical Behavior of Irradiated Metals and Alloys}, pages
  41--52, 1983.

\bibitem{Zinkle2013a}
S.~J. Zinkle, A.~M{\"{o}}slang, T.~Muroga, and H.~Tanigawa.
\newblock {Multimodal options for materials research to advance the basis for
  fusion energy in the ITER era}.
\newblock {\em Nuclear Fusion}, 53(10):13, oct 2013.

\bibitem{Getto2016}
E.~Getto, K.~Sun, A.~M. Monterrosa, Z.~Jiao, M.~J. Hackett, and G.~S. Was.
\newblock {Void swelling and microstructure evolution at very high damage level
  in self-ion irradiated ferritic-martensitic steels}.
\newblock {\em Journal of Nuclear Materials}, 480:159--176, nov 2016.

\bibitem{Dubuisson1992}
Philippe Dubuisson, A~Maillard, Christophe Delalande, Didier Gilbon, and
  Jean-Louis Seran.
\newblock The effect of phosphorus on the radiation induced microstructure of
  stabilized austenitic stainless steels.
\newblock Technical report, CEA Centre d'Etudes Nucleaires de Saclay, 1990.

\bibitem{Holmes1969}
J~J Holmes.
\newblock {Irradiation-induced swelling in nickel alloys}.
\newblock In {\em Battelle-Northwest, Richland, Wash.}, 1969.

\bibitem{VanDenBosch2013}
J.~{Van Den Bosch}, O.~Anderoglu, R.~Dickerson, M.~Hartl, P.~Dickerson, J.~A.
  Aguiar, P.~Hosemann, M.~B. Toloczko, and S.~A. Maloy.
\newblock {SANS and TEM of ferritic-martensitic steel T91 irradiated in FFTF up
  to 184 dpa at 413 °C}.
\newblock {\em Journal of Nuclear Materials}, 440(1-3):91--97, sep 2013.

\bibitem{Wolfer1984}
W.~G. Wolfer.
\newblock {Advances in void swelling and helium bubble physics}.
\newblock {\em Journal of Nuclear Materials}, 122(1-3):367--378, may 1984.

\bibitem{Russell1971}
K.~C. Russell.
\newblock {Nucleation of voids in irradiated metals}.
\newblock {\em Acta Metallurgica}, 19(8):753--758, aug 1971.

\bibitem{Okita2005}
T.~Okita, W.~G. Wolfer, F.~A. Garner, and N.~Sekimura.
\newblock {Effects of titanium additions to austenitic ternary alloys on
  microstructural evolution and void swelling}.
\newblock {\em Philosophical Magazine}, 85(18):2033--2048, jun 2005.

\bibitem{zinkle2017}
S.~J. Zinkle, J.~L. Boutard, D.~T. Hoelzer, A.~Kimura, R.~Lindau, G.~R. Odette,
  M.~Rieth, L.~Tan, and H.~Tanigawa.
\newblock {Development of next generation tempered and ODS reduced activation
  ferritic/martensitic steels for fusion energy applications}.
\newblock {\em Nuclear Fusion}, 57(9):17, jun 2017.

\bibitem{Dubuisson1993}
P.~Dubuisson, D.~Gilbon, and J.~L. S{\'{e}}ran.
\newblock {Microstructural evolution of ferritic-martensitic steels irradiated
  in the fast breeder reactor Ph{\'{e}}nix}.
\newblock {\em Journal of Nuclear Materials}, 205(C):178--189, oct 1993.

\bibitem{Singh1993}
B.~N. Singh and S.~J. Zinkle.
\newblock {Defect accumulation in pure fcc metals in the transient regime: a
  review}.
\newblock {\em Journal of Nuclear Materials}, 206(2-3):212--229, nov 1993.

\bibitem{Dai2020}
Y.~Dai, G.R. Odette, and T.~Yamamoto.
\newblock {The Effects of Helium in Irradiated Structural Alloys}.
\newblock In {\em Comprehensive Nuclear Materials}, pages 186--234. Elsevier,
  jan 2020.

\bibitem{li2019radiation}
Shi-Hao Li, Jing-Ting Li, and Wei-Zhong Han.
\newblock Radiation-induced helium bubbles in metals.
\newblock {\em Materials}, 12(7):1036, 2019.

\bibitem{farrell1980experimental}
K~Farrell.
\newblock Experimental effects of helium on cavity formation during
  irradiation—a review.
\newblock {\em Radiation Effects}, 53(3-4):175--194, 1980.

\bibitem{gigax2016radiation}
JG~Gigax, Tianyi Chen, Hyosim Kim, Jing Wang, LM~Price, Eda Aydogan,
  Stuart~Andrew Maloy, DK~Schreiber, MB~Toloczko, FA~Garner, et~al.
\newblock Radiation response of alloy t91 at damage levels up to 1000 peak dpa.
\newblock {\em Journal of Nuclear Materials}, 482:257--265, 2016.

\end{thebibliography}

\bibliographystyle{unsrt}

\end{document}